\documentclass[12pt, preprint]{aastex}
\newcommand{\cl}{\ensuremath{\mathcal{C}_{L}}}

\begin{document}

\title{Host Galaxies Catalog Used in LIGO Searches for 
Compact Binary Coalescence Events}
\author{Ravi Kumar Kopparapu\altaffilmark{1,2,*}, Chad Hanna\altaffilmark{2},
 Vicky Kalogera\altaffilmark{3}, Richard O'Shaughnessy\altaffilmark{3,*},
Gabriela Gonz\'alez\altaffilmark{2}, Patrick R. Brady\altaffilmark{4},
 Stephen Fairhurst\altaffilmark{4,5,6}}
\altaffiltext{1}{Center for Computation and Technology, Louisiana State
 University, Baton Rouge, LA  70803, USA}
\altaffiltext{2}{Department of Physics and Astronomy, Louisiana State
 University, Baton Rouge, LA  70803, USA} 
\altaffiltext{3}{Department of Physics and Astronomy, Northwestern University,
2145 Sheridan Road, Evanston, IL 60208, USA} 
\altaffiltext{4}{Department of Physics, University of Wisconsin-Milwaukee,
P.O. Box 413, Milwaukee, WI 53201, USA} 
\altaffiltext{5}{School of Physics and Astronomy, Cardiff University,
Cardiff, CF2 3YB, United Kingdom.}
\altaffiltext{6}{LIGO Laboratory, California Institute of Technology, Pasadena,
CA 91125, USA} 
\altaffiltext{*}{At present: Center for Gravitational Wave Physics,
104 Davey lab, Pennsylvania State University, University Park,
 PA - 16802-6300, USA}



\begin{abstract}

 An up-to-date catalog of nearby galaxies  considered as hosts of 
binary compact objects is provided with complete information about sky position,
 distance, extinction-corrected blue luminosity and error estimates.  
 With our current understanding of binary evolution, rates of
formation and coalescence for binary compact objects scale with massive-star
formation
 and hence the (extinction-corrected) blue luminosity of host galaxies.
  Coalescence events in
binary compact objects are among the most promising gravitational-wave sources
for ground-based gravitational-wave detectors such as LIGO.  Our catalog
and associated error estimates are important for the
interpretation of analyses, carried out for LIGO, to constrain the rates of
compact binary coalescence, given an astrophysical population model for the
sources considered.  We discuss how the notion of effective distance, created
to account for the antenna pattern of a gravitational-wave detector, must be
used in conjunction with our catalog.
We note that the catalog provided can be used on other astronomical
analysis of populations that scale with galaxy blue luminosity. 

 
\end{abstract}


\section{INTRODUCTION}
\label{introduction}


Compact binary coalescence (CBC) events, such as neutron star
or black hole mergers, are one of the primary gravitational-wave sources for
ground-based interferometers such as LIGO\footnote{www.ligo.org}.
 LIGO's third (S3, Oct 31 2003 - Jan 9 2004) and fourth (S4, Feb 22 2005 - 
Mar 23 2005)
 science runs have reached 
significant extragalactic distances \citep{LIGOS3S4iul}
 into the nearby Universe.   
Especially for massive compact binaries whose components are black holes,  
the range extended beyond the Virgo 
Cluster. To interpret the searches for signals from compact binary coalescence
in the LIGO data sets,
it is necessary to use information about 
putative binary compact object populations in the known nearby galaxies, as well
as how the population scales at larger distances. 
 The nearby galaxy catalog discussed here represents the
 distribution of such extragalactic populations, and the procedures described 
are used for LIGO data analysis, such as assigning an astrophysically 
meaningful upper limits given non-detection. An accurate upper-limit 
that correctly incorporates  our best
information about galaxy distributions requires a model of the nearby
 overdense region because the current LIGO network's range 
probes this overdensity.


Binary compact objects are usually
produced from the evolution of massive stellar binaries.  Since
short-lived, massive stars emit more blue light than all other stars
in a galaxy combined, blue light is a well-known tracer of star formation in
general and the birthrate of these massive stars in particular.
Given the short lifetimes of the known Milky Way double compact object
population and the slow rate of change in star formation expected
in nearby and distant galaxies,  \citet{Phinney:1991ei} has argued
that a galaxy's blue luminosity should linearly scale with its
compact binary coalescence rate.

%
%
The sensitivity of  LIGO  to compact binary coalescence signals
depends on the distance and sky position of the coalescence event and
 therefore, 
the distribution of known nearby galaxies in blue luminosity and in 
space is the minimum information needed to properly interpret searches of
 the LIGO data sets. 

It is possible that compact binary populations that are not related to
regions of star formation may exist in the Universe.
A mass, metallicity and morphology dependent star
formation history may be needed to account for these populations\footnote{
\cite{Lipunov1995} adopts a mass normalization to derive
their event rate of 1/year within 50 Mpc using an older version of
 Tully's catalog, whereas
we use a blue-light normalization and the up-to-date Tully catalog
(\S\ref{section2}). 
Their study also differs from ours because we consider issues like antenna
 pattern of the detector and completeness corrections, which they ignore.}.
Nevertheless, the work described here is limited
to the blue-light luminosity as a tracer of the compact binary population.

 The contribution of elliptical galaxies to the merger rates
is potentially significant beyond the Virgo cluster \citep{deFreitasPacheco}, 
whereas their blue luminosity is not representative of their putative compact 
binary populations.
 However, at large distances,  the fractional  blue luminosity
produced in ellipticals is  about $10 \%$
  \citep{Driver2007}, and at  short 
distances the contribution is negligible because there are fewer
ellipticals in the nearby local universe.  \cite{deFreitasPacheco}
 conclude that
the event rate for an elliptical galaxy with the same blue luminosity as a
 spiral
galaxy is a factor of five times larger on average.
We conclude that LIGO rate upper limits derived from the catalog presented here would change by less than a factor of 1.5 due to a correction for elliptical galaxies.

Our blue light census will also implicitly not account for any potential
contribution from globular clusters to the compact binary coalescence
rate of the nearby universe.  
\citet{Phinney:1991ei} has argued that the contribution of globular clusters to
 double neutron star mergers in the Galaxy would not exceed 10 \% of the
 coalescence rate due to the Galactic field.
On the other hand it has been argued that the contribution of globular clusters 
to binary black hole coalescence may be very significant
 \citep[see, e.g.,][]{PZMcM,clusters-2005}.
However, these cluster contributions are expected to become significant
 at distances
 beyond the Virgo cluster, where a more significant fraction of
 ellipticals with large globular-cluster systems will eventually enter LIGO's
 detection volume.


We have used mostly publicly available astronomical catalogs of 
galaxies to compile a catalog used in the S3/S4/S5 (fifth science run\footnote{
http://lhocds.ligo-wa.caltech.edu/scirun/S5/}, Nov 4 2005 - present)
LIGO data 
set analyses. We discuss the methodology used to compile this 
galaxy catalog and briefly describe how this information feeds into
LIGO rate estimates.   
In \S\ref{section2}, we describe all the elements involved in compiling the
 galaxy catalog and assessing the relevant errors and uncertainties.
In \S\ref{section3}, we derive a correction factor to account for
 incompleteness in the catalog guided also by the blue-light volume density
 estimated from the Sloan Digital Sky Survey and earlier surveys. In 
\S\ref{section4}, we
discuss how the corrected catalog and resulting blue light distribution as a
function of distance is used to bound the rate of compact binary coalescence
using data from the recent LIGO science runs. If the maximum distance
to which a search could detect a compact binary coalescence is known,
then the expected number of detectable events can be derived.  Some
concluding remarks are made in \S\ref{section5}.


\section{COMPILATION OF GALAXY CATALOG}
\label{section2}

We have compiled a catalog\@\footnote{%
http://www.lsc-group.phys.uwm.edu/cgi-bin/cvs/viewcvs.cgi/lalapps/src/ \\
inspiral/inspsrcs100Mpc.errors?cvsroot=lscsoft},
the \emph{compact binary coalescence galaxy} catalog
or CBCG-catalog, of nearby galaxies which
could host compact binary systems. For each galaxy out to 100 Mpc, the
catalog provides the equatorial coordinates, distance to the galaxy,
and the blue luminosity corrected for absorption.  Estimates of the
systematic errors on distance and luminosity are also provided.

The CBCG-catalog is compiled from information provided in the following
four astronomical catalogs: (i) the Hubble Space 
Telescope (HST) key project catalog used to measure the Hubble constant 
\citep{Freedman:2001}, (ii) Mateo's dwarf galaxies of the local group 
catalog \citep{Mateo:1998}, (iii) the HyperLeda (LEDA) database of galaxies
\citep{LEDA}, and (iv) an
updated version of the Tully Nearby Galaxy Catalog \citep{Tully:TPC}.

When combining these catalogs, distances and luminosities reported in
the HST, Mateo and Tully catalogs were generally adopted over those in
the LEDA catalog. This is because these catalogs use accurate 
distance determination methods compared to LEDA.
 Nevertheless, LEDA served as the baseline for
comparisons in the range 10-100 Mpc since it is the most complete.

\subsection{DISTANCES} 
\label{distance_section}
One of the primary objectives of 
the HST key project was to discover Cepheid variables (stars which
have periodic variations in brightness) in several nearby spiral
galaxies and measure their distances accurately using the 
period-luminosity relation for Cepheids. 
Cepheid distance determination to nearby galaxies is one of 
the most important and accurate primary distance indicators.
The distance information from the HST key project is considered to be
the most accurate in the CBCG-catalog; there are 30 galaxies in our
catalog for which we adopt distances from the HST key project.   

Mateo's review \citep{Mateo:1998} of properties of the 
dwarf galaxies in the Local Group provides distance and 
luminosity information for each galaxy considered. 
Since the parameters in this catalog were derived from focused studies on each 
individual galaxy,
we consider it the most accurate next to the HST measurements for nearby 
galaxies. Moreover it has reasonably comprehensive information on the Local 
Group's dwarf galaxies; there are 18 sources in the CBCG-catalog which adopt
distances (and luminosities) from Mateo's compilation.

It becomes increasingly difficult to use primary distance estimators
like Cepheid stars in more distant galaxies. Therefore secondary distance 
methods are used to measure larger distances.
Tully's catalog has up to three types of distances for each source:
(i){\it Quality distance} ($D_\mathrm{Q}$) is based on either Cepheid 
measurements, surface brightness fluctuations, 
or the tip of the red giant branch. There are 409 galaxies 
with such a distance in the CBCG-catalog. 
(ii) {\it HI luminosity-line-width} distances ($D_\mathrm{HI}$) are 
obtained from the Tully-Fisher relation, where the maximum rotational 
velocity of a galaxy (measured by the Doppler broadening of the 
21-cm radio emission line of neutral hydrogen) is correlated with 
the luminosity (in B, R, I and H bands) to find the distances. There
are 553 galaxies in the catalog with such a distance. 
(iii) {\it Model distance} ($D_\mathrm{M}$) is derived from an evolved
dynamical mass model that translates galaxy radial velocities into 
distances. This model is an update of the least action model 
described by \cite{Shaya95} and takes into account the deviations 
from a perfect Hubble flow due to a spherically symmetric 
distribution of mass centered on the Virgo Cluster.  All
galaxies have  a calculated model distance. 
Whenever 
available, $D_\mathrm{Q}$ distances are the most preferred due to 
their smaller uncertainties, then the  $D_\mathrm{HI}$ 
 followed by $D_\mathrm{M}$.

The remaining galaxies come from LEDA which 
 does not provide distances explicitly, but instead provides
 measured radial velocities corrected
for in-fall of the Local Group towards the Virgo cluster 
($v_\mathrm{vir}$).  We obtain
the LEDA distance ($D_\mathrm{L}$) using Hubble's law
 with the Hubble constant $H_\mathrm{0} = 73\,\mbox{km}\,\mbox{s}^{-1}\,\mbox{Mpc}^{-1}$  reported by 
\cite{Spergel}.
 Although corrections to the recessional velocity
were made, this method of calculating distances is still highly
uncertain. Hence, we use Hubble's law to evaluate the 
distances only to the galaxies for which $v_\mathrm{vir} \ge 500$ 
km/s (7Mpc) and peculiar velocities are expected to be
more of a perturbation.

\begin{figure}[!hbp|t]
\centering
\epsscale{1}
\plotone{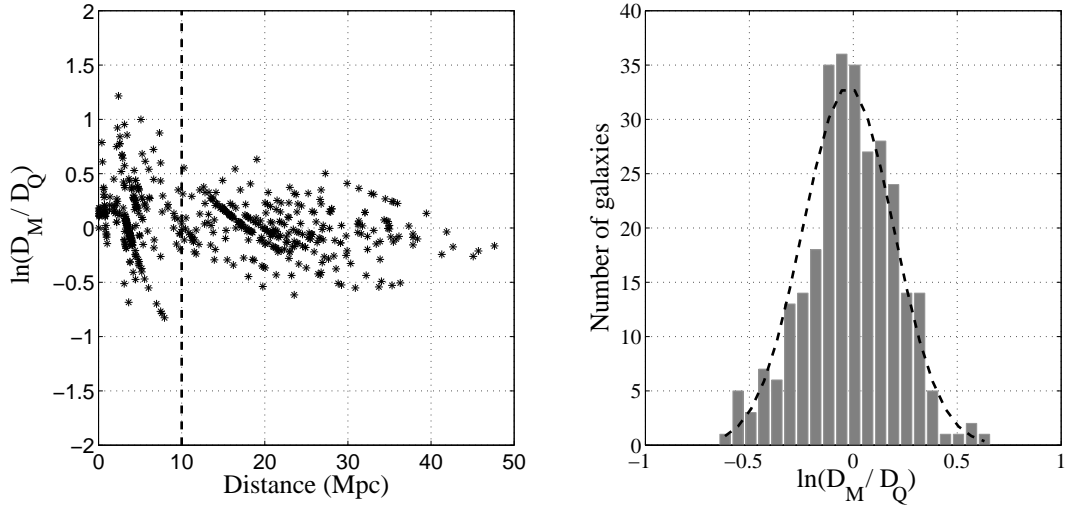}
\hfill \caption{\label{TullyDistanceErrors}In order to obtain 
reasonable estimates for Tully's model distances we compare 
galaxies that have values for both.  We only consider galaxies beyond
10 Mpc since model distances and LEDA distances are not reliable below
this value.  All galaxies below 10 Mpc have better distance estimates.
The Tully quality distance has roughly a $0.1$ logarithmic error.  
 The best fit 
Gaussian for $\ln[D_M / D_Q]$ implies a fractional error $\sigma$ of 
$0.24$ in log.  Subtracting these uncertainties in quadrature gives an error of 
$0.22$ for Tully model distances.}
\end{figure}

The error in a distance depends strongly on the 
method used to measure that distance. The HST sources, 
though a small contribution to the galaxy catalog, have the smallest 
errors ($< 10 \%$) \citep{Freedman:2001}. 
 The three different distance methods in 
Tully's catalog have different errors. $D_\mathrm{Q}$
also has a low error ($10 \%$) followed by the 
$D_\mathrm{HI}$ 
($20 \%$). To obtain an estimate for the errors of
$D_\mathrm{M}$, we compare them with $D_\mathrm{Q}$ 
 for the set of galaxies that have both types of distance estimates.
The best fit Gaussian 
(see Fig.~\ref{TullyDistanceErrors}) to
the logarithm of fractional errors has
a one sigma width of  $0.24$ which when subtracted in quadrature
with $D_\mathrm{Q}$ error gives, 
$0.22$ distance error associated with $D_\mathrm{M}$.

\begin{figure}[!hbp|t]
\centering
\epsscale{1}
\plotone{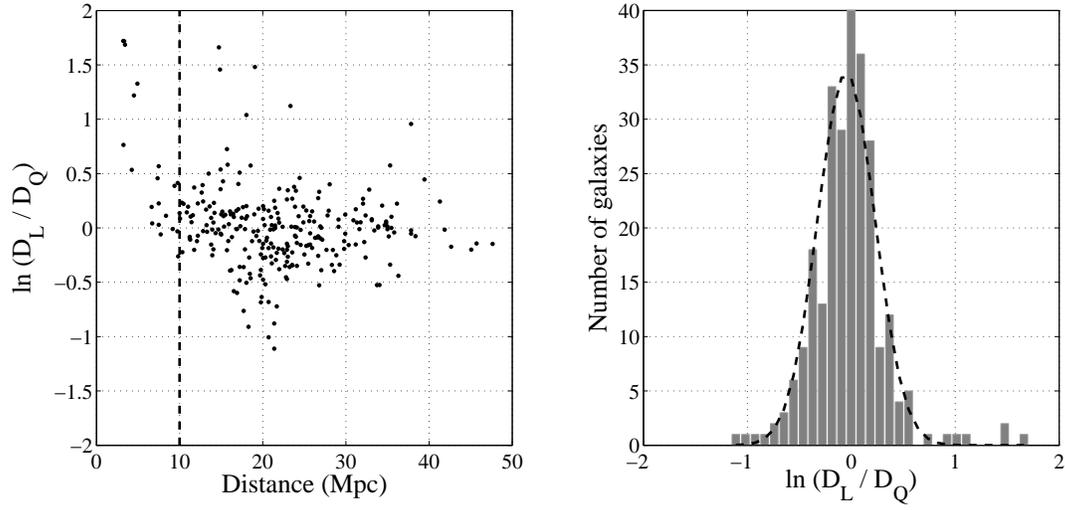}
\hfill \caption{\label{DistanceErrors}
Fractional error analysis as in Fig.~\ref{TullyDistanceErrors} for LEDA
distances.   By comparing the fractional error between LEDA
distances and Tully we obtain a $\sim$ $0.25$ log distance error for LEDA.}
\end{figure}

Because errors in $v_\mathrm{vir}$ are not given in LEDA, we follow a
similar procedure to find LEDA distance errors, $D_\mathrm{L}$.  We
compare the calculated $D_\mathrm{L}$ with $D_\mathrm{Q}$ for galaxies
in both catalogs to obtain uncertainty estimates in $D_\mathrm{L}$.  The
plot in Fig.~\ref{DistanceErrors} shows the best fit Gaussian to the
logarithm of fractional errors with a one sigma width of $0.27$ which,
 subtracted in
quadrature with $D_\mathrm{Q}$ distance errors, gives a total distance error
$0.25$.\@\footnote{For searches of the S3 and S4 LIGO data
\citep{LIGOS3S4iul}, with smaller ranges a more conservative uncertainty
of 40\% was used for LEDA distances.} 
\clearpage
\clearpage
\clearpage

\subsection{BLUE LUMINOSITIES} 

%

The distribution of binary compact objects 
in the nearby universe is expected to follow the 
star formation in the universe and a measure of star formation is the 
blue luminosity of galaxies corrected for dust extinction and 
reddening \citep{Phinney:1991ei}. Hence, for each galaxy, we calculate 
the blue luminosity $L_\mathrm{B}$ from the
absolute blue magnitude of the galaxy $M_\mathrm{B}$ 
(corrected for internal and Galactic extinctions). 
For convenience, blue luminosity is provided in units
of $L_\mathrm{10} \equiv 10^{10} L_\mathrm{B, \odot}$, where 
$L_\mathrm{B, \odot} = 2.16 \times 10^{33}$ ergs/s is the blue
solar luminosity derived from the blue solar magnitude
 $M_\mathrm{B, \odot} = 5.48$
\citep{Binney-Tremaine}.
We do not consider galaxies with luminosities 
less than $10^{-3} L_\mathrm{10}$ because they do not contribute significantly
to the total luminosity -- see \S\ref{section3}.

\begin{figure}[!hbp|t]
\centering
\epsscale{1}
\plotone{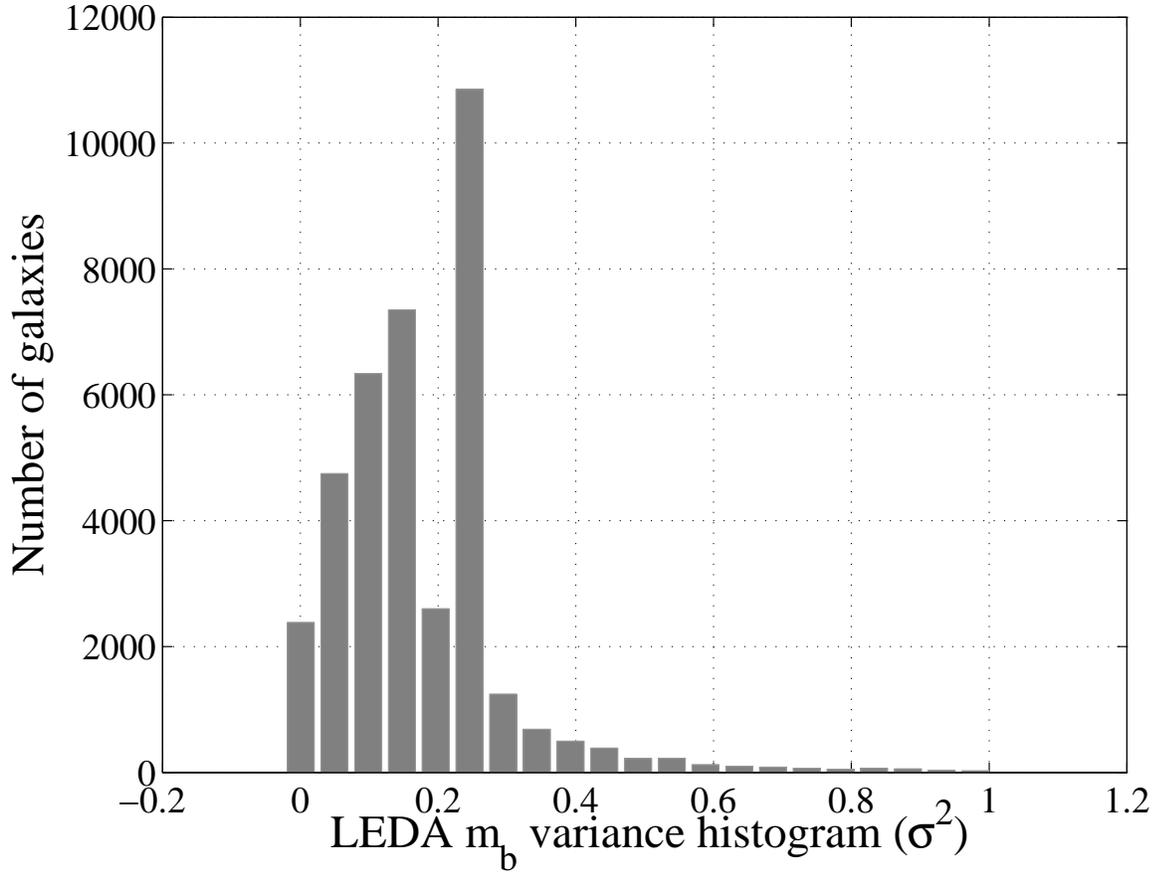}
\caption{\label{LEDAmagerrCorrected}LEDA provides uncertainties in apparent
magnitudes.  The histogram above shows the $m_b$ variance distribution 
for each LEDA galaxy.   The RMS error is 0.42.}
\end{figure}

The Mateo, Tully and LEDA catalogs provide information on 
apparent B-magnitudes 
corrected for extinction. The galaxies in the HST key project catalog
have only distance information, so for those we extract
the corresponding apparent magnitude values ($m_\mathrm{B}$, 
corrected for internal and Galactic extinction) in the B-band from the 
Tully catalog to find $M_\mathrm{B}$. 
Table~\ref{table1} summarizes relevant properties of each of these
catalogs and the fraction of the total luminosity within 100 Mpc that
each contributes. 
\clearpage
\begin{table*}[!h]
\caption{Summary information about the four astronomical catalogs used to develop
the CBCG-catalog. We report the number of galaxies for which the catalog was the
primary reference and fraction of the total CBCG-catalog blue luminosity
accounted for by those galaxies.}
\begin{center}
\label{table1}
\begin{tabular}{@{} lllllr @{}}
\hline
\hline
& Catalog&\# of galaxies& ${\rm L}_{10}$& Fractional luminosity & Reference \\
&&& $(10^{10} L_\mathrm{B,\odot})$&& \\
\hline
(i) & HST&30&57.3 &0.1$\%$&\citep{Freedman:2001}\\
(ii) & Mateo&18&0.4 &$<$0.001$\%$&\citep{Mateo:1998}\\
(iii) & Tully&1968&2390 &5.3$\%$& \citep{Tully:TPC}\\
(iv) & LEDA&36741&42969.4 &94.6$\%$&\citep{LEDA}\\
\hline
\hline
& Total&38757&45417.1 & $100.0$\%  \\
\hline
\end{tabular}
\end{center}
\end{table*}
\clearpage
The LEDA database quotes uncertainties in apparent magnitude. 
Figure~\ref{LEDAmagerrCorrected} shows the distribution of
LEDA assigned apparent magnitude variances for the galaxies in the 
CBCG-catalog. The RMS error is $\Delta m_B = 0.42$.
Galaxies from Tully's catalog have a smaller observational error
$\Delta m_\mathrm{B} = 0.30$ \citep{Tully:TPC}.
\clearpage
\clearpage

\section{COMPLETENESS}
\label{section3}
\begin{figure}[!hbp|t]
\centering
\epsscale{1}
\plotone{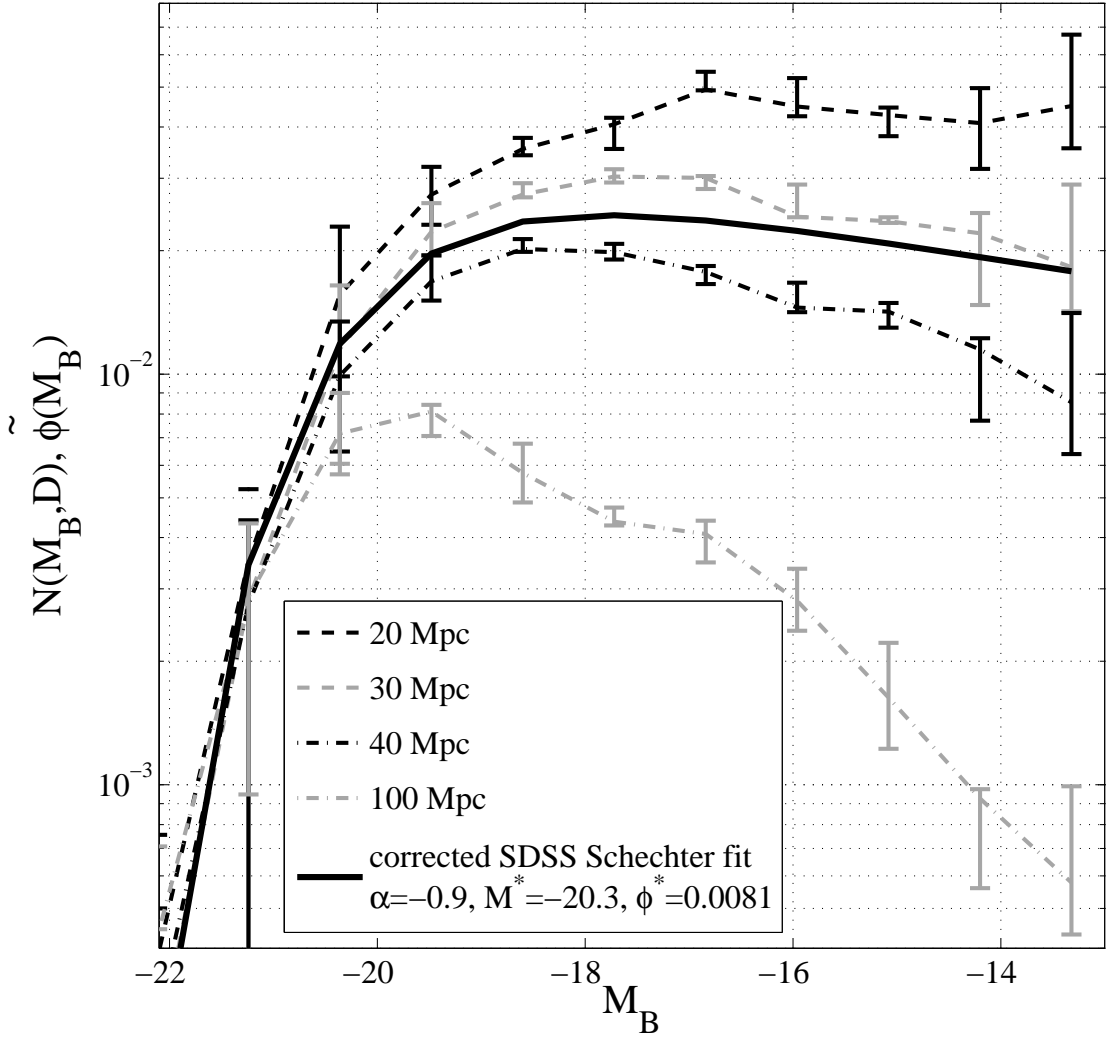}
\hfill \caption{\label{lum_function} The luminosity function of CBCG
catalog at various distances
 (dashed and dot-dashed lines) and a Schechter function fit (solid line)
given in Eq. (\ref{phi_MB}) based on \cite{Blanton}.
We compensate for the incompleteness of the CBCG-catalog by applying an
upward correction to 
the
luminosity bins that are below the Schechter function fit (solid line),
according to Eqs. (\ref{Ltot}) and (\ref{Lcorr}). 
Error bars are found by sliding the magnitudes
of each galaxy according to the mean errors and recomputing the luminosity 
function.}
\end{figure}

Observations of faint galaxies  are difficult even in the nearby 
universe and lead to systematic incompleteness in galaxy 
catalogs. Studies of galaxy luminosity
functions can provide insight into how many galaxies are missing from
a catalog (and hence the corresponding blue luminosity). Using the
CBCG-catalog, we can generate a luminosity function $N(L,D)$ which is
the number of galaxies with luminosities within a luminosity bin from
$L$ to $L+\Delta L$ normalized to the spherical volume within radius
D. Specifically, we write
\begin{equation} 
N(L,D) \Delta L = \left(\frac{3}{4\pi D^3}\right) \,\left[\sum_\mathrm{j} l_\mathrm{j}\right] 
\label{N_LD}
\end{equation} 
where
\begin{eqnarray}
\nonumber
l_\mathrm{j} =
\left\{
\begin{array}{ll}
1 & \hspace{1.cm} {\rm if }\,\,(L < L_\mathrm{j} < L+\Delta L) \textrm{ and } (D_j < D) \nonumber \\
0 & \hspace{1.cm} {\rm otherwise}
\end{array}
\right.
\end{eqnarray}
and the sum over $j$ runs through all the galaxies in the catalog. The
quantities $L_\mathrm{j}$ and $D_\mathrm{j}$ are the luminosity and
distance of each galaxy. Similarly we can compute the luminosity
function in terms of blue absolute magnitudes as a function of
distance $N(M_\mathrm{B},D)$. The dashed and dot-dashed lines in
 Fig.~\ref{lum_function} show
several realizations of $N(M_\mathrm{B},D)$ for different distances
$D$ plotted as a function of $M_\mathrm{B}$. 

To estimate the degree of incompleteness in the CBCG-catalog,
we use an analytical Schechter galaxy luminosity function  \citep{Schechter}
\begin{equation}
\phi (L) dL = \phi^{*} \biggl(\frac{L}{L^{*}}\biggr)^{\alpha} \exp\biggl(\frac{-L}{L^{*}}\biggr)  d\left(\frac{L}{L^{*}}\right)
\label{phi_LUM}
\end{equation}
where $\phi (L) dL$ is the number density (number of galaxies per 
unit volume) within the luminosity interval $L$ and $L + dL$, 
$L^{*}$ is the luminosity at which the number of galaxies begins to fall off 
exponentially, $\alpha$ is a parameter 
which determines the slope at the faint end of the luminosity function, and
$\phi^*$ is a normalization constant. 
In terms of (blue) absolute magnitudes, $M_\mathrm{B}$, the Schechter function 
becomes
%
\begin{eqnarray}
\label{schecter}
\tilde {\phi}(M_\mathrm{B}) dM_\mathrm{B}= 0.92\, {\phi^{*}} 
  \exp\left[{-10^{-0.4(M_\mathrm{B} - {M_\mathrm{B}^{*}})}}\right]
  \left[ 10^{-0.4(M_\mathrm{B} - {M_\mathrm{B}^{*}})} \right]^{\alpha + 1} \, dM_\mathrm{B} \; .
\label{phi_MB}
\end{eqnarray}
%

To estimate the total luminosity function, we use results from
the Sloan Digital Sky Survey (SDSS) as reported by \cite{Blanton}.
Although the SDSS sky coverage is inadequate in RA and DEC, it provides
excellent coverage throughout our desired distance and
beyond.  We therefore use the green luminosity function Schechter fit
given in Table 2. of \cite{Blanton} and convert it into blue band using the expression
given in Table 2. of \cite{BlantonRoweis}.
 Adopting a Hubble constant value of  $73\,\mbox{km}\,\mbox{s}^{-1}\,\mbox{Mpc}^{-1}$
 \citep{Spergel} and correcting for reddening,\@\footnote{We correct 
the value of $M_\mathrm{B}^{*}$ to be consistent with the reddening
 correction described in \S\ref{cubic_law}}
 the Schechter parameters are
$(M_\mathrm{B}^{*}, \tilde{\phi^{*}}, \alpha) = (-20.3, 0.0081, -0.9)$.
The solid line in Fig.~\ref{lum_function} 
shows the Schechter function $\tilde \phi(M_\mathrm{B})$ derived from these
 values. Since this function is obtained from deep surveys, it does not 
account for the local over-density of blue light coming primarily from the
Virgo cluster. For distances up about to 30 Mpc, the CBCG-catalog's 
luminosity function $N(M_\mathrm{B}, D)$ exceeds
 $\tilde \phi(M_\mathrm{B})$.

We can now derive a completeness correction that arises at the faint end
beyond about 30 Mpc, where the Schechter function exceeds the catalog
$N(M_\mathrm{B}, D)$. We integrate the CBCG-galaxy-catalog luminosity
function $N(L,D)$ over $L$ and subtract it from the Schechter fit as a
function of distance. Hence, the total 
corrected cumulative luminosity $L_{\mathrm{total}}$ within a volume of radius $D$  
is given by
\begin{equation}
L_{\mathrm{total}}(D) = L_{\mathrm{CBCG}}(D) + L_{\mathrm{corr}}(D)
\label{Ltot}
\end{equation}
where
\begin{eqnarray}
L_{\mathrm{CBCG}}(D) &=& \int_{0}^{D} dD' \sum_\mathrm{j}L_\mathrm{j}
            \delta(D'-D_\mathrm{j}) \label{Lcat} \\
L_{\mathrm{corr}}(D) &=& \frac{4\pi}{3}D^3\int_{L_\mathrm{min}}^{L_\mathrm{max}}L\,dL
\,\,\Theta \left[\phi(L)-N(L,D)\right]\,\, \left[ \phi(L)-N(L,D) \right] \; .
\label{Lcorr}
\end{eqnarray}
Here, the index $j$ runs through all galaxies in the catalog, $\delta$ is the
 Dirac delta function, $\Theta$ is the step function and $\phi(L)$ is the 
adopted Schechter function (distance independent) assumed to represent the
 complete luminosity distribution. We note that $L_\mathrm{max} = 52.481 
~L_\mathrm{10}$ ($M_\mathrm{B} = -23.83$) is the maximum luminosity 
in the CBCG-catalog and we choose
 $L_\mathrm{min} = 10^{-3}L_\mathrm{10}$ ($M_\mathrm{B} = -12.98$) 
 because luminosities below this  value do not contribute
significantly to the net luminosity.
 The quantity $L_{\mathrm{CBCG}}$ in Eqs.~(\ref{Ltot}) and (\ref{Lcat}) is the
uncorrected cumulative luminosity from the CBCG-catalog; the
quantity $L_{\mathrm{corr}}$ is the completeness correction.  Note that the completeness 
correction term is always zero or positive regardless of the choice of
Schechter function.  

In Fig.~\ref{cum_lum}, we show the cumulative blue luminosity as a
function of distance as obtained directly from the CBCG-catalog
(solid line) as well as with the completeness correction applied
(dashed line). It is evident that the correction becomes significant
at distances in excess of about 40Mpc. 

\clearpage
\clearpage

\subsection{Comparison with other results}
\label{cubic_law}

\begin{figure}[!hbp|t]
\centering
\epsscale{1}
\plotone{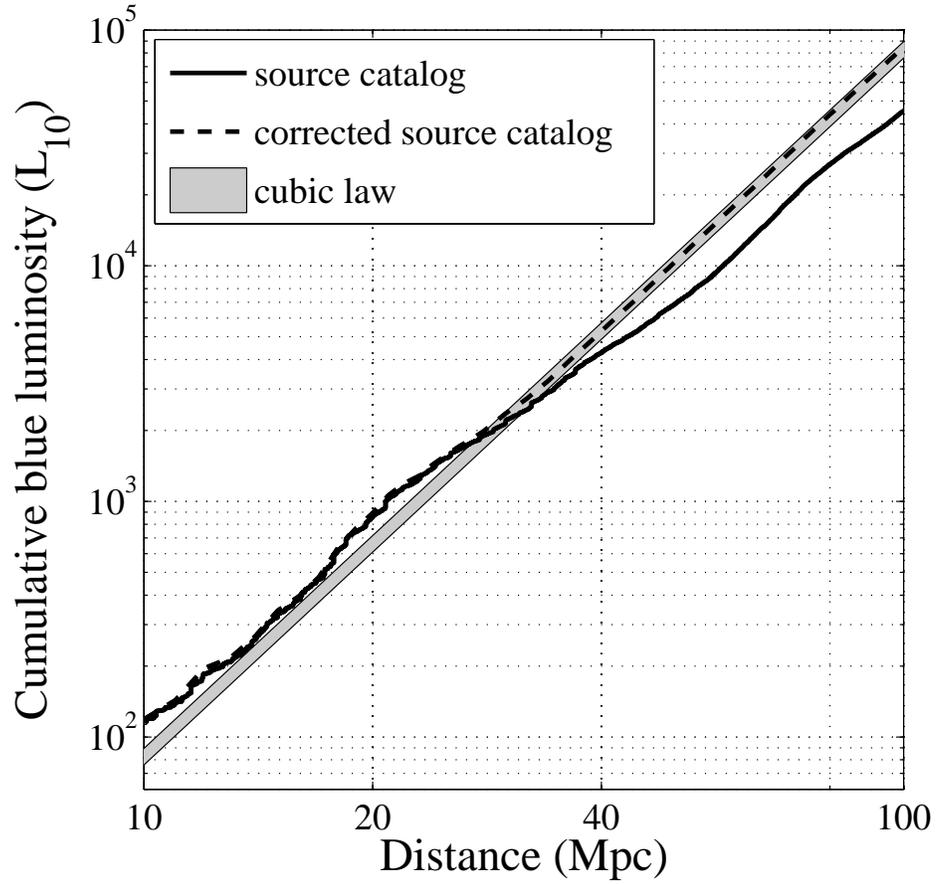}
\hfill \caption{\label{cum_lum} Cumulative luminosity  as a function of 
distance from CBCG-catalog uncorrected for incompleteness 
(solid line), corrected for incompleteness 
(dashed line) and the cubic extrapolation from the assumed constant blue
 luminosity density corrected for extinction (gray-shaded region).}
\end{figure}

To compare our method of correcting for completeness with other methods,
 we consider the direct
computation of a reddening corrected luminosity density based on
\cite{Blanton} which could be used at large distances. 
We adopt a blue luminosity density of $(1.98\pm 0.16) \times 10^{-2} 
L_\mathrm{10}/$Mpc$^{3}$ calculated as follows:
\begin{itemize}
\item The blue luminosity density, in terms of 
  blue absolute magnitudes per cubic Mpc, is  $-14.98$ locally (redshift $z = 0$ )
  and $-15.17$ for $z = 0.1$ [Table 10 \cite{Blanton}].
 This is for a standard cosmology with 
  $\Omega_\mathrm{M} = 0.3$ and $\Omega_\mathrm{\Lambda} = 0.7$. 
We use $z=0.1$ so that the results will be valid for advanced
detectors. 
\item We convert the $z = 0.1$ blue magnitude density (-15.17) to
 luminosity units 
 $1.33 \times 10^{-2} L_\mathrm{10}/$Mpc$^{3}$ 
  and assign systematic errors ($\simeq 10 \%$) associated with the
 photometry 
 to obtain a luminosity density of  
  $(1.33 \pm 0.13) \times 10^{-2} L_\mathrm{10}/$Mpc$^{3}$.
\item  We also correct for processing of blue light and 
  re-emission in the infrared (IR) following \cite{Phinney:1991ei}
 and \cite{Kalogera:2000dz}.
 We use the analysis of \cite{Saunders}, upward correct by $30 \%$ their 
far IR ($40 \mu m - 100 \mu m$) luminosity density
to account 
  for emission down to $12 \mu m$ \citep{Kalogera:2000dz},
 and convert to $L_\mathrm{10}$ 
  to obtain an IR luminosity density of 
  $L_\mathrm{IR} = (0.65 \pm 0.1) \times 10^{-2} 
  L_\mathrm{10}/$Mpc$^{3}$.
\item  Adding both luminosity densities above and accounting for
  the errors, we obtain a blue light luminosity density corrected for
 extinction equal to $(1.98 \pm 0.16) \times 10^{-2} 
  L_\mathrm{10}/$Mpc$^{3}$ 
  \end{itemize}
%
We use this blue luminosity density and its uncertainty and plot the implied
cumulative blue luminosity as a function of distance (cubic dependence)
in Fig.~\ref{cum_lum} (gray-shaded region). This uniform density
distribution agrees well with the completeness corrected luminosity
given above.

We can compare our results for the cumulative blue luminosity as a
function of distance to similar results obtained by \cite{Nutzman:2004},
especially their Figure 1. The results for the uncorrected catalog 
agree qualitatively.  However,
the catalog described here is more up-to-date compared to the one
compiled by \cite{Nutzman:2004} by virtue of the updates to LEDA and by
the inclusion of the current Tully catalog. The incompleteness
correction derived here is also more physically and empirically
motivated than the one constructed in that earlier paper.  We note that
the cumulative luminosity shown as the dashed line in their Figure 1 is
too low by a factor of $4\pi/3$  due to a numerical error. Additionally,
their luminosity density is $\sim 25 \%$ lower than ours resulting from our
use of the more recent results presented by \cite{Blanton}.

\clearpage
\clearpage


\section{COMPACT BINARY COALESCENCE RATE ESTIMATES}
\label{section4}

For neutron star binaries, the observed binary pulsar sample can be
used to predict the coalescence rate $\mathcal{R}_{\mathrm MW}$ in the Milky
Way~\citep{Kim:2004,Kim:2006}. The coalescence rate within a sphere of
radius $D$ is then simply given by 
\begin{equation}
R = \mathcal{R}_{\mathrm MW} 
\left(\frac{L_{\mathrm{total}}(D)}{L_{\mathrm MW}}\right) \; 
\end{equation}
where $L_\mathrm{total} (D)$ is the total blue luminosity within a distance $D$ 
and $L_\mathrm{MW}$ is the blue luminosity of the Milky way,
 $1.7 L_\mathrm{10}$ \citep{Kalogera:2000dz}.
If the rate $R$ of a binary neutron star coalescence could be measured
directly, it would provide an independent estimate of the rate of
coalescence per unit of blue luminosity. Together these two
measurements would deepen our understanding of stellar and binary
evolution. Furthermore, the current understanding of binary evolution
and compact object formation leads us to anticipate the formation of
black hole binaries that will merge within a Hubble time
\citep{Belczynski:2002, Belczynski:2007}. Experiments like LIGO will 
provide a direct measure of the compact binary coalescence rate and
will impose constraints on the theoretical models of stellar evolution
and compact binary formation. 

\subsection{Rate estimates and systematic errors in gravitational-wave
searches}
In its simplest form, the rate estimate derived from a
gravitational-wave experiment will take the form
\begin{equation}
{\mathcal R} =  \frac{\textrm{constant}}{ T \, \cl}
\end{equation}
where the constant depends on the precise outcome of the search and
the statistical method used in arriving at the rate estimate, $\cl$ is
the cumulative blue luminosity {\it observable} within the search's
sensitivity volume measured in $L_{10}$, and $T$ is the time analyzed
in years. In general the sensitivity volume is a complicated function
which depends on the instrument and the gravitational waveforms
searched for.  Here, we focus on the influence of the host galaxy
properties and the distribution of blue light with distance.

The gravitational-wave signal from a compact binary inspiral depends on
a large number of parameters. It is convenient to split these parameters
into two types for our discussion. Of particular interest here are the
parameters which determine the location and orientation of the binary.
We denote these collectively as $\vec{\lambda} := \{ D, \alpha, \delta,
\iota, \psi, t\}$, that is the distance to the binary, its Right
Ascension and declination, inclination angle relative to the line of
sight, polarization angle of the waves, and the time when the binary is
observed, respectively.  Other parameters, including the masses and the spins, are
denoted $\vec{\mu}$. Recognizing that the spatial luminosity
distribution can be written as 
\begin{equation}
L(\alpha,\delta,D) = \sum_j L_j \, \delta(\alpha_j - \alpha ) 
\, \delta( \delta_j - \delta ) \, \delta( D_j - D ) \; ,
\label{e:spatlum}
\end{equation}
we write the cumulative luminosity as
\begin{equation}
{\cl} = \int L(\alpha, \delta, D) \, 
p(\textrm{detection} | \vec{\mu}, \vec{\lambda}) \, 
p( \vec{\mu} ) \, p(\iota ) \,  p(\psi) \,  p(t) \, 
d\vec{\mu} \, d\vec{\lambda} 
\label{e:cumlum}
\end{equation}
Assuming that binary coalescences are uniformly distributed in time, and
their orientation is random, we take the corresponding prior
probabilities:  $p(\iota) = \sin (\iota )/2$, $p(t)
= 1/\textrm{day}$, and $p(\psi ) = 1/2\pi$.

Systematic errors associated with the derived rate esimates are
naturally associated with the errors in cumulative luminosity $\cl$.
The two most relevant errors in the galaxy catalog are in apparent
magnitude $m_B$ and distance $D$.  Sky positions are known so precisely
that small errors in RA and DEC do not change the detection probability
of a particular binary in any significant way; for this reason, such
errors are not included in the LIGO analyses~\citep{LIGOS3S4iul}. The
errors induced on the spatial luminosity function in
Eq.~(\ref{e:spatlum}) take the form \citep{Fairhurst}

\begin{equation}
  [L + \Delta L](\alpha,\delta,D) = \sum_j L_j \, 10^{- 0.4 \Delta m_{Bj}}
  \left(1 + \frac{\Delta D_j}{D_j} \right)^2
  \delta(\alpha_j - \alpha ) \,
  \delta( \delta_j - \delta ) \,
  \delta( D_j + \Delta D_j  - D ) \; .
  \label{e:spatlumerror}
\end{equation}

\subsection{A simplified model for estimating expected event rates}

The sensitivity of a search for gravitational waves from compact binary
coalescence is determined primarily by the amplitude of the waves at the
detector.  For a non-spinning binary 
(i.e., the spins of each 
 compact  object are much smaller than their general-relativistic maximum
 value of $m_\mathrm{i}^2$) with given $\vec{\mu}$, the
amplitude is inversely proportional to the \textit{effective distance}
${D}_{\rm eff}$ defined as \citep{Findchirp}
\begin{equation}
\label{effDist}
{D}_{\rm eff} =  \frac{D}{\sqrt{F_+^2 (1 + \cos^{2} \iota)^2 / 4 + F_{\times}^2
\cos^{2} \iota}}  
\end{equation}
where $D$ is the physical distance to the binary, $F_\mathrm{+}$ and 
$F_\mathrm{\times}$ are the response amplitudes of each polarization
at the detector which depend upon the location of the binary system
\citep{Anderson}:
\begin{eqnarray}
F_\mathrm{+} = -\frac{1}{2}(1+\cos^{2} \theta) \cos 2 \phi \cos 2 \psi
- \cos \theta \cos 2 \phi \sin 2 \psi \, \\ 
F_\mathrm{\times} = \frac{1}{2}(1+\cos^{2} \theta) \cos 2 \phi \sin 2 \psi
- \cos \theta \sin 2 \phi \cos 2 \psi \;  .
\end{eqnarray}
Here $\theta$ and $\phi$ are the spherical co-ordinates of the source
defined with respect to the detector and, as before, $\iota$ and $\psi$
are the inclination and polarization angles.  Since $\theta$ and $\phi$
are detector dependent, the effective distance is different for
geographically separated detectors that are not perfectly aligned and,
for a fixed source location, changes as the Earth rotates through a
sidereal day.  Additionally, the effective distance is always at least
as large as the physical distance.

\begin{figure*}[!hbp|t]
\centering
\epsscale{1}
\plotone{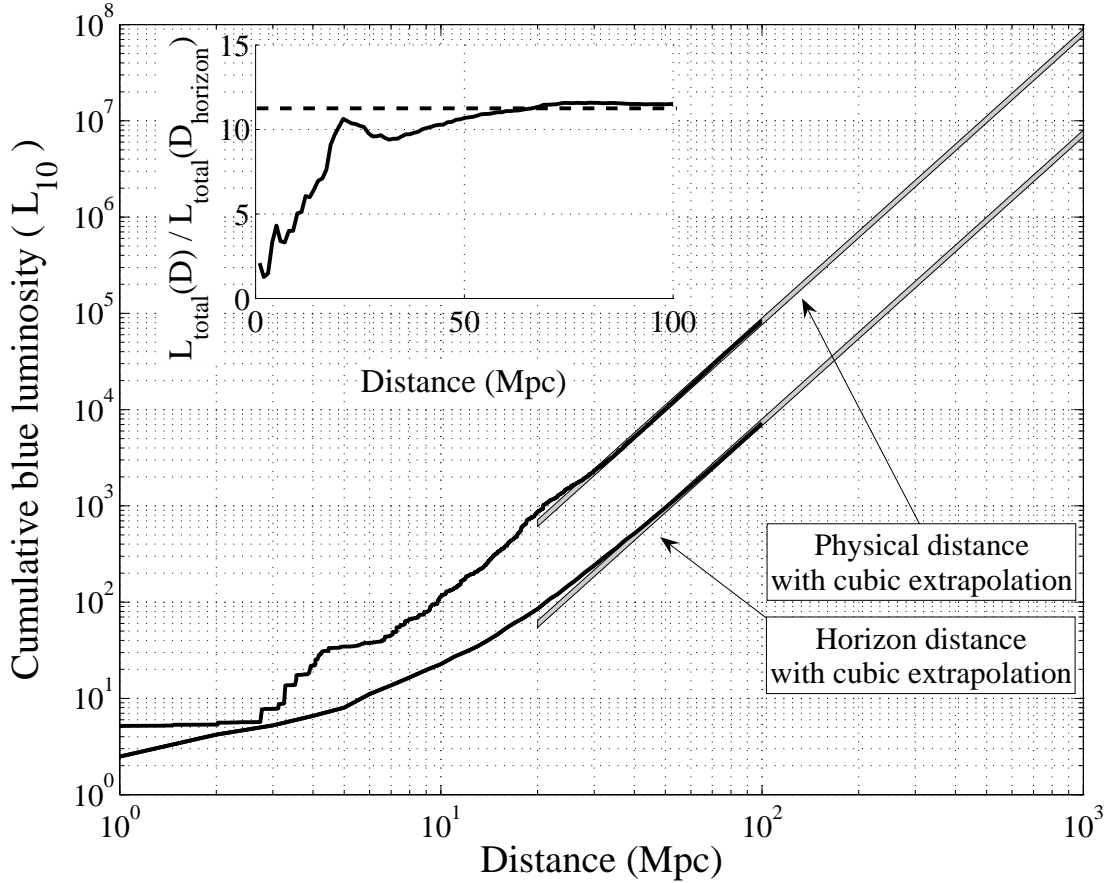}
\hfill \caption{\label{cumlum_effdist}
Cumulative luminosity as a function of physical distance (top line)
and horizon distance (bottom line).
The horizon distance $D_{\mathrm{horizon}}$ is defined as the physical
distance to an optimally oriented and located binary system that
would be detected with a signal-to-noise ratio of 8.  (Instrumental
sensitivity range is sometimes quoted in terms of the radius of a sphere
with the same volume as the non-uniform region probed by the
instrument, this sensitivity range $D_s$ is related to the horizon
distance by $D_s\simeq D_{\mathrm{horizon}}/\sqrt{5}$.
The gray shaded lines are
cubic extrapolations ($\S\ref{section3}$) derived for
both cases.  Given a LIGO horizon distance one can immediately get the
cumulative blue luminosity from the bottom curve.
To obtain an approximate rate upper limit
one could calculate $\mathcal{R}_{90\%} \, [ \,\mathrm{yr}^{-1}L_{10}^{-1} ]
 = 2.3/(\mathcal{C}_{L} \times T)$ where $\mathcal{C}_{L}$ is taken
from this plot at a given range in horizon distance. {\it Inset:} Ratio
of the cumulative luminosity for the physical and horizon distance from
the completeness corrected CBCG-catalog illustrates
the non-uniform distribution at smaller ranges ($< 20$ Mpc) and
asymptotes to the expected uniform distribution ratio (dashed line) for
 larger distances.}
\end{figure*}

For simplicity in understanding the sensitivity of gravitational-wave
searches, consider the case in which $\vec{\mu}$ is fixed, i.e.
$p(\vec{\mu}) = \delta(\vec{\mu} - \hat{\mu})$.  For example, these
might be the parameters appropriate to a neutron star binary.  
The sensitivity of a detector is given by its horizon distance, which
is defined as the maximum effective distance that a neutron star binary
system can be detected at signal-to-noise ratio of 8.
Consider a
search which can perfectly detect these binaries if they have an
effective distance $D_{\mathrm{eff}} < D_{\mathrm{horizon}}$ at a
particular detector. Then 
\begin{equation}
p(\textrm{detection} | \hat{\mu}, \vec{\lambda}) = 
\Theta( D_{\mathrm{eff}}(\vec{\lambda}) < D_{\mathrm{horizon}} ) \,
\end{equation}
and we can write 
\begin{equation}
{\cl}(D_{\mathrm{horizon}}) = \int L(\alpha, \delta, D)  \, 
\Theta( D_{\mathrm{eff}}(\vec{\lambda}) < D_{\mathrm{horizon}} ) \,
p(\iota ) \, p(\psi) \, p(t) \,  
d\vec{\lambda} \, .
\label{e:cumlumdist}
\end{equation}
Thus, the cumulative blue luminosity accessible to such a detector is
the blue luminosity within an effective distance sphere of
radius $D_{\mathrm{horizon}}$, averaged over the time of day and
possible orientations of the binary. The lower curve in 
Fig.~\ref{cumlum_effdist} shows $\cl(D_{\mathrm{horizon}})$. 
Figure \ref{cumlum_effdist} also illustrates the significant difference
between the cumulative luminosity $\cl (D_{\mathrm{eff}})$ and total
luminosity $L_{\mathrm{total}}(D)$ at a given distance.  If galaxies are
distributed uniformly in space the ratio between these is $\simeq
11.2$ ; this is the factor by which the detection rate would be 
reduced and arises purely from the LIGO detector  
response, averaged over all possible source orientations with respect to the 
detector.

When estimating the rate based on gravitational-wave observations,
one can marginalize over uncertainties \citep{Fairhurst} in the galaxies'
distances and apparent magnitudes.  Specifically, by making use of the 
modified spatial distribution function Eq.~(\ref{e:spatlumerror}) and the 
distributions for $\Delta D_j$ and $\Delta m_{Bj}$ reported here, we can 
obtain a probability distribution for the cumulative luminosity
$p(\cl | \Delta m_{Bj}, \Delta D_j)$ from Eq.~(\ref{e:cumlum}).  For each
value of the cumulative luminosity, a probability distribution 
$p(R | \cl)$ for the event rate can be calculated.  Finally, the rate is
marginalized over errors in the galaxy catalog by computing
\begin{equation}\label{eq:marginalize}
  p(R) = \int d \cl \,\, p(\cl | \Delta m_{Bj}, \Delta D_j) \,\,  p(R | \cl) 
  \, .
\end{equation}
This distribution is then used to obtain a rate interval or upper limit on 
the occurrence of binary coalescences in the unverse.

\begin{figure*}[!hbp|t]
\centering
\epsscale{1}
\plotone{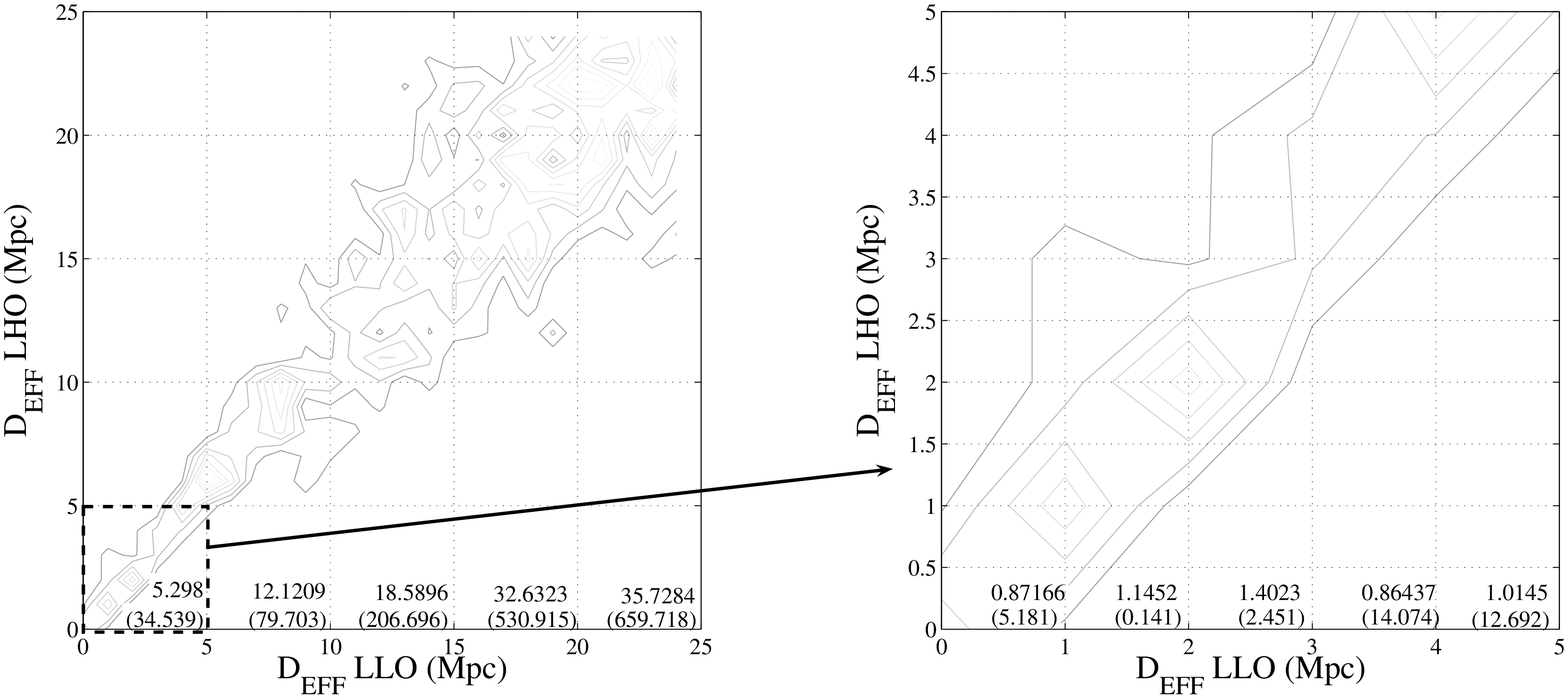}
\hfill \caption{\label{LumCont}
Luminosity contours per effective distance bin in the two
LIGO sites. The 
effective distance to a source in one galaxy is different between
  the two detectors, 
changes as a function of the sidereal day and also on the orientation of the
particular source.  Since the effective
distance is always larger than the real distance the luminosity 
available within a given effective distance bin is 
considerably smaller than the luminosity within the physical 
distance bin.  The upper horizontal numbers refer to the luminosity per
bin in effective distance.  The parenthetical lower numbers refer
to the luminosity per physical distance bin.  It is also possible 
to have a systematically different luminosity between detectors as is
indicated in the right panel zoom of the first 5 Mpc.  The available
luminosity within 5 Mpc (mostly from Andromeda) is slightly 
better located for LLO and therefore stretches  
the contours to higher effective distances for LHO. LIGO rate 
upper limits for searches with limited range thus depend 
on the non-uniformity of the Local Group.}
\end{figure*}

While this approach provides a reasonable estimate of the observable
blue light luminosity in a single detector, it does not provide the
whole story. For example, the $16^\circ$ difference in latitude between
the LIGO Observatories in Hanford, Washington and Livingston, Louisiana,
implies the $\cl(D_{\mathrm{horizon}})$ depends on the site used.
Figure \ref{LumCont} shows two-dimensional contours of this function. 

%

\clearpage

\clearpage
\clearpage

Based on the galaxy catalog presented in this article, the cumulative 
blue luminosity $\cl$, measured in $L_{10}$, accessible to a search with a 
given horizon distance sensitivity can be derived from Fig.~\ref{cumlum_effdist}
and is tabulated in Table~\ref{t:instrumental-rates}. 
We can combine the calculated cumulative blue luminosity 
with estimates of ${\mathcal R}$, the rate of binary 
mergers per $L_{10}$, to estimate the number of compact binary merger 
events $N$ detectable in a given LIGO search 
with an observation time $T$:
\begin{equation}
N = 
 10^{-3} 
\left(\frac{\mathcal R}{L_{10}^{-1} \textrm{ Myr}^{-1}} \right)
\left(\frac{\cl}{10^3 L_{10}}\right)
\left(\frac{T}{\textrm{ yr}}\right)
\end{equation}
If the horizon distance of a search is larger than 50 Mpc, we can use the following approximation, from a cubic law:  
\begin{equation}
N \approx
7.4 \times 10^{-3} 
\left(\frac{\mathcal R}{L_{10}^{-1} \textrm{ Myr}^{-1}} \right)
\left( \frac{D_{\mathrm{horizon}}}{100 \textrm{ Mpc}} \right)^3
\left(\frac{T}{\textrm{ yr}}\right)
\label{e:nevents-approx}
\end{equation}

Estimated rates of binary neutron star (BNS) mergers in our Galaxy are
based on the observed sample of binary pulsars. The rates depend on the
Galactic distribution of compact objects. In \cite{Kalogera:2004tn}, the
most recent reference estimating rates, the most likely Galactic rate
for their reference model 6 is $83\,\mbox{Myr}^{-1}$, with a 95\%
confidence interval $17-292 \, \mbox{Myr}^{-1}$. The most likely rates for
all the models used in \cite{Kalogera:2004tn} are in the range
$4-220\,\mbox{Myr}^{-1}$ for the Milky Way.\@\footnote{The rates quoted here
are in units of rate per Milky Way per Myr; to get the rate per
$L_{10}$, we divide by $1.7$ which is the estimated blue luminosity of
the Milky way in $L_{10}$ units, assuming the blue absolute magnitude of
the Milky Way to be $-20.11$ \citep{Kalogera:2000dz}.}

For the 4km LIGO detectors currently operating, 
$D_{\mathrm{horizon}}
\approx 30 \textrm{ Mpc}$ for BNS.  
Thus, the predicted number of BNS events is in the range $N_6 \approx 2  - 
30\times 10^{-3} \textrm{ yr}^{-1}$
with the most likely number being $N_6 \approx 1/(100 \textrm{ yr})$ 
[we use the subscript $_6$ to indicate these rates use reference model 6 from 
\cite{Kalogera:2004tn}]. 
A search that reaches twice the distance (such as enhanced LIGO),
yields a most likely rate $
N_6 \approx 1/(10 \textrm{ yr})$.
And a search that would be 15 times more sensitive to coalescences of binary
 systems than the current LIGO detectors (such as 
Advanced LIGO) would yield a most likely rate of 
$N_6 \approx 40.0 \textrm{ yr}^{-1} $. 
\clearpage
\begin{table}
\caption{\label{t:instrumental-rates}
Table showing the cumulative blue luminosity
$\cl(D_{\mathrm{horizon}})$ accessible to a search
with horizon distance $D_{\mathrm{horizon}}$ given in the first column.
For $D_{\mathrm{horizon}} > 100 \textrm{ Mpc}$, the cumulative blue
luminosity accessible to a search is given approximately by
$C_L(D_{\mathrm{horizon}}) \approx 7.4 \times 10^3 \,\, (D_{\mathrm{horizon}} 
/ 100 \textrm{Mpc})^3$. }
\begin{center}
\begin{tabular}{|c|c|c|c}
\hline
$D_{\mathrm{horizon}}$ (Mpc) & $\cl(D_{\mathrm{horizon}})$ ($L_{10}$)  \\
\hline
$10$ & $23$ \\
$20$ & $85$  \\
$30$ & $240$ \\
$50$ & $953$  \\
$100$ & $7200$ \\
$200$ & $59200$  \\
$300$ & $200000$ \\
$500$ & $926000$  \\
\hline
\end{tabular}
\end{center}
\end{table}
\clearpage

\section{CONCLUSION}
\label{section5}

Whether one wishes to compute expected detection rates for LIGO
searches, or to interpret LIGO searches as rate upper limits (or eventually
detection rates), we require at the simplest level accurate accounting of
the total observable blue luminosity $\cl$. 
As mentioned in the previous sections, a galaxy catalog complete with
sky positions and distances is important for first generation LIGO detectors 
because the blue luminosity is not uniformly distributed 
in the sky within the search range.  An upper limit which takes in to
account the most up-to-date information on galaxy distribution can be 
obtained by accurately modeling the local overdense region.
For searches with ranges well beyond current sensitivity the universe
is uniform and rate estimates depend primarily on accurate blue 
luminosity densities corrected for reddening.  We have 
introduced a method to bridge the gap between the well known nearby
galaxy distribution and the expected long range distribution through
a completeness correction based on SDSS luminosity functions 
\citep{Blanton}.

This paper provides the most up to date accounting of nearby galaxies
within 100Mpc as well as errors in the apparent magnitude (corrected for
reddening) and distance and demonstrates how the errors propagate 
into rate calculations.  
 Astrophysical errors are a significant contribution to the
eventual systematic error associated with coalescence rate upper limits
 \citep{Fairhurst} and must be  included.
This paper provides a survey of the asymptotic and local uncertainty.
Motivated  
by the use of effective distance to account for the antenna pattern  
of the LIGO detectors, we demonstrate the need to compute the average  
blue light luminosity within a given effective distance sphere.
  For ranges within 50Mpc there is a nontrivial relationship
between cumulative blue luminosity within an effective distance sphere and
within a physical distance sphere.  Beyond 50Mpc the relationship is well
behaved leading to the simple scaling for the number of detected events $N$ 
given in Eq.(\ref{e:nevents-approx}).  We would like to point that the
 catalog
 provided can also be used on other astronomical analysis of populations
 that scale with galaxy blue luminosity, such as the local
 Type II supernova rate or the rate of nearby SGR bursts that show up
 as short GRBs.

We provide sufficient  description of our methods for others
to apply new rate models to future LIGO data.  Although this catalog 
will serve as a reference for current and future LIGO data analysis, we
look forward to future work that may transcend the simple blue light
rate normalization that we have discussed.  One way to go beyond blue light
rate normalization, (necessary to ascertain the degree to which old stars 
contribute to present day mergers) is with multiband photometry of nearby 
galaxies which can  reconstruct their mass, 
morphology and metallicity dependent star formation
history.  With this information in hand LIGO detections could be applied more
stringently to assess the relative contribution that progenitors of different
ages provide to the present day merger rate. 


\acknowledgements

We would like to thank B. Tully for generously providing his most 
up to date nearby galaxies catalog in the preparation of this work. We
 acknowledge the usage of the HyperLeda database (http://leda.univ-lyon1.fr).
 We also thank  P. Nutzman and the members of the LIGO Scientific 
Collaboration Compact-Binary-Coalescence group for many
 insightful discussions.
This work has been supported in part by 
NSF grants PHY-0200852, PHY-0353111, PHY 03-26281, PHY 06-00953, PHY 06-53462,
PHY-0355289, AST-0407070, a David and Lucile
Packard Foundation Fellowship in Science and Engineering (VK),  a
Cottrell Scholar Award from the Research Corporation (PRB),
the Royal Society (SF) and 
Center for Computation and Technology (RKK). This work was also supported by
 the Center
for Gravitational Wave Physics, which is supported by the National
Science Foundation under cooperative agreement PHY 01-14375.
LIGO was constructed by the California Institute of Technology and
Massachusetts Institute of Technology with funding from the National
Science Foundation and operates under cooperative agreement PHY-0107417.
This paper has LIGO Document Number LIGO-P070065-00-Z.

\end{document}